\documentclass[10pt, fleqn]{extarticle}

\usepackage{amsmath}
\usepackage{amsfonts}
\usepackage{amssymb}
\usepackage{graphicx}

\usepackage{epstopdf}
\usepackage{fancyhdr}
\usepackage{tikz}
\usepackage{caption}

\usepackage{float}
\usepackage{subcaption}
\usepackage[italicdiff]{physics}

\usepackage[margin = 0.8in, a4paper]{geometry}
\usepackage{listings}
\usepackage{mathtools}
\usepackage{cancel}

\usepackage{hyperref}
\usepackage{url}
\usepackage{tensor}
\usepackage{amsthm}

\usepackage{mdframed}
\usepackage{epigraph}
\usepackage[style = numeric-comp, backend = biber]{biblatex}
\usepackage[skins, breakable]{tcolorbox}

\usepackage{todonotes}
\usepackage{rotating}
\usepackage{multirow}
\usepackage{array}

\usepackage{enumitem}
\usepackage{zref-clever}
\usepackage{csquotes}
\usepackage{lettrine}

\usepackage{setspace}
\author{Robbert W.\ Scholtens \and Marcello Seri \and Holger Waalkens \and Rien van de Weygaert}
\title{Exploring the CMB in Anisotropic Universes}

\mdfdefinestyle{myStyle}{
	linecolor = black,
	linewidth = 1pt,
	frametitleaboveskip = -1em,
	innertopmargin = -.5em,
	shadow = true,
	shadowsize = 2.5pt,
	shadowcolor = gray,
	skipbelow = 5pt,
	roundcorner = 0pt
}

\numberwithin{equation}{section}
\numberwithin{figure}{section}
\numberwithin{table}{section}

\theoremstyle{definition}

\tcbuselibrary{theorems}

\zcsetup{
	lang = english,
	abbrev = true,
	capfirst = true}

\zcRefTypeSetup{section}{Name-sg = \S\!\!}
\zcRefTypeSetup{appendix}{Name-sg = \S\!\!}
\zcRefTypeSetup{notation}{Name-sg = Notation}

\tcbset{
	colback = white,
	colframe = black,
	fonttitle = \bfseries\color{black},
	enhanced,
	breakable,
	boxed title style = {colback = white, colframe = black},
	attach boxed title to top left = {yshift = -4mm, xshift = 4mm},
	top = 5mm,
	description delimiters none,
	separator sign dash,
	sharp corners,
	drop shadow = gray
}

\newtcbtheorem[number within = section]{boxthm}{Theorem}{label type = theorem}{thm}
\newtcbtheorem[use counter from = boxthm]{boxdef}{Definition}{label type = definition}{def}
\newtcbtheorem[use counter from = boxthm]{boxlem}{Lemma}{label type = lemma}{lem}
\newtcbtheorem[use counter from = boxthm]{boxprop}{Proposition}{label type = proposition}{prop}
\newtcbtheorem[use counter from = boxthm]{boxrem}{Remark}{label type = remark}{rem}
\newtcbtheorem[use counter from = boxthm]{boxalgo}{Algorithm}{label type = algorithm}{algo}
\newtcbtheorem[use counter from = boxthm]{boxexample}{Example}{label type = example}{ex}
\newtcbtheorem[use counter from = boxthm]{boxnotation}{Notation}{label type = notation}{not}

\newtcolorbox{proofbox}{title = \normalfont{\textit{Proof.}}}

\newcommand{\tb}[1]{\boldsymbol{#1}}

\newcommand{\Lder}[1]{\mathcal{L}_{#1}}

\newcommand{\contr}[2]{\left\langle#1\,;#2\right\rangle}

\renewcommand{\qedsymbol}{$\blacksquare$}

\begin{document}
	\maketitle
	\noindent\emph{Author affiliations and contact details at end of document.}
	\\
	\begin{abstract}
		\noindent In recent years, there have been increasing challenges to the cosmological principle, based on new observations of e.g. supernovae and the cosmic bulk flow. As a result, the cosmological community is speaking their concern for the cosmological principle, and from which scales onwards it should apply. In this context, there is a desire to understand more fully the properties and signatures of cosmologies not obeying the cosmological principle. In this article, we let go of the demand of cosmic isotropy, and instead assume only spatial homogeneity in our cosmological models. We follow the results of our previous works [see citations in the list of references], and here bring these together into one unified picture, with the goal of describing the signature(s) of anisotropy in anisotropic cosmological models. We first introduce the Bianchi models---a particular instance of spatially homogeneous cosmologies---and show that a metric can be constructed for them when an appropriate collection of desired Killing vector fields is supplied. Then, we give the perturbations of the Friedmann equations in such Bianchi models, in the Newtonian gauge, derived using much the same methodology as applicable to the FLRW models. We show these can be combined into one characteristic partial differential equation. Finally, we use this equation in order to simulate the CMB of a toy Bianchi V example and produce its power spectrum. We close with a discussion, and suggestions for further research.
	\end{abstract}
	
	\section{Introduction}
	Our modern understanding of cosmology, encapsulated in $\Lambda$CDM, is based on the assumption of the \emph{cosmological principle.} This states the universe is spatially homogeneous and isotropic when considered over sufficiently large scales. Roughly speaking, the cosmological principle means that every location in the universe is equivalent, as well as every direction, agreeing well with the Copernican principle that we are not at a preferred location in space. Mathematically, the cosmological principle allows us to drastically reduce the potential spacetime metrics valid at the aforementioned sufficiently large scales: indeed, we obtain the well-known \emph{Robertson-Walker} family of spacetimes, which lead to the flat, open, and closed FLRW cosmological models.
	
	Over recent years, a growing list of observations have lead to the realization that the universe may be somewhat less perfect than expressed by the cosmological principle and that we need to consider and study slight modifications to the assumption of perfect homogeneity and isotropy on the very largest scales. Examples of observations that have led to questions on the cosmological principle are for instance, supernovae \cite{secrest_test_2021, secrest_colloquium_2025, wagenveld_cosmic_2023}, large-scale bulk flows \cite{watkins_analysing_2023, watkins_origins_2025}, anisotropy in observations of the Hubble parameter \cite{colin_probing_2011, sah_anisotropy_2025, luongo_larger_2022}, and observations of gamma ray bursts in the observed CMB dipole direction \cite{luongo_exploring_2025}---for an excellent, recent review, see \cite{aluri_is_2023}. At recently convened platforms, cosmologists have spoken their concern for the cosmological principle, or at least that the scale on which we should consider its validity is higher than previously conceived.
	
	Our work focuses on letting go of isotropy as demanded by the cosmological principle, and investigating spacetimes which are only spatially homogeneous. Specifically, we wish to investigate CMBs in such universes, so as to better understand possible signatures of anisotropy in, for instance, observations of the CMB as made by the \emph{WMAP} \cite{bennett_nine-year_2013} and \emph{Planck} \cite{planck_collaboration_planck_2020} missions. Being the initial condition for the development of large-scale structures in the universe, furthering the understanding of the CMB and how it may be influenced by cosmic anisotropy is of paramount importance to modern cosmology. Here we take the opportunity to bring together the results of our works \cite{scholtens_fundamental_2025, scholtens_fundamental_2026, scholtens_fundamental_2026-1}, and provide a unified picture of these works. This includes describing spatially homogeneous universes mathematically, and writing down perturbation equations in such universes. We also apply our methodology to a Bianchi V universe, to obtain concrete CMB maps.
	
	In \zcref{sec:theory} we briefly introduce the mathematical structure of spatially homogeneous, anisotropic universes, discussing the mathematical structure that remains. In particular, we narrow down to our intended field of research, namely the \emph{Bianchi models.} In \zcref{sec:perturbations}, then, we find the perturbation equations of the energy density $\rho$, pressure $p$, momentum flux density $\tb{q}$, and anisotropic stress $\tb{\pi}$, in terms of our chosen geometries, assuming we have chosen scalar perturbations in the Newtonian gauge. We use those for the energy density and pressure to create one, wave-like equation that governs scalar perturbations. Finally, in \zcref{sec:bianchi-v-cmb}, we introduce our methodology for finding a CMB map of a Bianchi V universe example. We close off this article with a small discussion, and prospects for further research.
	
	\section{Describing anisotropy---the Bianchi models}\label{sec:theory}
	Letting go of isotropy means that we expand the class of possible spacetimes beyond the Robertson-Walker family. In this section, we discuss the spacetimes that can occur when maintaining only spatial homogeneity, letting go of istropy. We first consider them from a topological point of view, showing that they still admit a bona fide cosmic time, and then from the geometric point of view, choosing a frame that best encapsulates them.
	
	Let us consider first spatial homgeneity topologically. Mathematically, spatial homogeneity means that there exist three linearly independent, purely spatial Killing vector fields (KVFs) on spacetime $\{\tensor{\tb{\xi}}{_i}\}_{i=1}^3$ (so at each event $x$, the resulting vectors $\{\tensor{\tb{\xi}}{_i}|_x\}_{i=1}^3$ are linearly independent), which in turn generate the isometry group belonging to the metric of the relevant spacetime. Since the KVFs taken together form a Lie algebra, we can apply Frobenius' theorem in order to conclude that the orbits of the isometries generated by the KVFs taken together form a foliation of spacetime. Intuitively, every leaf of the foliation of spacetime consists of events that can be reached by applying an isometry---generated by the KVFs---to any other event. If this can be done uniquely, we say the action of the group of isometries is \emph{simply transitive}. Let us make the assumption that indeed the action is simply transitive; then the spatially homogeneous models that we are considering possess \emph{exactly} three KVFs as described above. These are known as \emph{Bianchi models,} and have been studied extensively since the 1950s and `60s \cite{taub_empty_1951, ellis_class_1969, ellis_class_1970, maccallum_cosmological_1973, ellis_bianchi_2006}. They owe their name to the Italian mathematician Luigi Bianchi, who showed at the start of the 20th century that the 3-dimensional Lie algebras can be classified \cite{bianchi_three-dimensional_2001}. Now the fact that we can label each leaf by some number $t$ shows the existence of a cosmic time function $t:\mathcal{M}\to\mathbb{R}$, like we have in the case of the Robertson-Walker spacetimes.
	
	The specific goal of our work \cite{scholtens_fundamental_2025} is as follows: when given a set of vector fields $\{\tensor{\tb{\xi}}{_i}\}_{i=1}^3$, to find a metric $\tb{g}$ such that these $\tb{\xi}$s are \emph{KVFs for that metric.} As we show constructively in that work, under mild assumptions on the $\tb{\xi}$s we can find a collection of vector fields $\{\tensor{\tb{X}}{_i}\}_{i=1}^3$ on our spacetime such that $\Lder{\tensor{\tb{\xi}}{_i}}\tensor{\tb{X}}{_j}=[\tensor{\tb{\xi}}{_i},\tensor{\tb{X}}{_j}]=\tb{0}$, \emph{expressing the $\tb{X}$s in the basis afforded to us by the $\tb{\xi}$s.} (See also tabulations in \cite{ryan_homogeneous_1975, stephani_exact_2003} for explicit expressions.) This vanishing commutator implies that (i) the $\tb{X}$s have a Lie algebra structure isomorphic to that of the $\tb{\xi}$s, and (ii) the duals to the $\tb{X}$s, say $\{\tensor{\tb{\omega}}{^i}\}_{i=1}^3$ so that $\contr{\tensor{\tb{\omega}}{^i}}{\tensor{\tb{X}}{_j}}=\tensor*{\delta}{^i_j}$, also vanish under Lie dragging by the same KVFs. By introducing $\tensor{\tb{\omega}}{^0}\equiv\dd{t}$, then, we find that the spacetime metric which has (at least) the $\tb{\xi}$s among its KVFs is expressible as
	\begin{equation}
		\dd{s}^2=\tensor{g}{_\alpha_\beta}(t)\,\tensor{\tb{\omega}}{^\alpha}\tensor{\tb{\omega}}{^\beta}.
	\end{equation}
	The essence here is that the $\tb{\omega}$s are in general not exact, have spatial dependence, but are independent of the cosmic time $t$, in the sense that $\Lder{\tensor{\tb{X}}{_0}}\tensor{\tb{\omega}}{^\mu}=\tb{0}$. So, in a spatially homogeneous spacetime, we are able to write the metric in a generally \emph{non-coordinate} (co-)frame, but such that the metric \emph{components} depend only on cosmic time $t$. This single dependence of the metric components transforms Einstein's field equations from partial to ordinary differential equations, grearly improving tractability, both analytically and computationally.
	
	Although work of a similar character has been conducted earlier \cite{taub_empty_1951, jantzen_dynamical_1979}, ours distinguishes itself by presenting the metric (co)frame not in a canonical coordinate system, but instead as linear combinations of the metric's (desired) KVFs. A key advantage of our approach is that, should some coordinate system have already been established, one does not need to perform an additional coordinate change in order to rewrite the metric into the form of the above. Instead, one uses our method to calculate the metric (co)frame directly from the KVFs. This may be advantageous in, for instance, experiments wherein some external coordinate system is preferred to coincide with experimental constraints, or simulations where the coordinates are based on computational convenience.
	
	The price that must be paid is in the interpretation of the tensor forms, since these now rely on non-coordinate frames, and some further geometric complexity. As a direct demonstration the connection symbols of the Levi-Civita connection are given in this frame by
	\begin{equation}\label{eq:levi-civita}
		\tensor{\Gamma}{^\sigma_\mu_\nu}=\tfrac{1}{2}\tensor{g}{^\alpha^\sigma}\left(2\,\tensor{g}{_\alpha_{(\mu,\nu)}}-\tensor{g}{_\mu_\nu_{,\alpha}}\right)+\tensor{g}{^\alpha^\sigma}\tensor{C}{^\beta_\alpha_{(\mu}}\tensor{g}{_{\nu)}_\beta}-\tfrac{1}{2}\tensor{C}{^\sigma_\mu_\nu},
	\end{equation}
	where comma notation indicates vector field application, and the $\tb{C}$ are the \emph{structure constants} of the frame: $[\tensor{\tb{X}}{_i},\tensor{\tb{X}}{_j}]=\tensor{C}{^a_i_j}\tensor{\tb{X}}{_a}$, with $\tensor{C}{^0_\mu_\nu}=\tensor{C}{^\sigma_0_\nu}\equiv0$. When working in a coordinate frame, the structure constants vanish identically and we are left with the well-known Christoffel symbols. The structure constant now being generally non-trivial impacts the connection, and resultantly all the curvature tensors that are built from it. Most importantly, though: all the \emph{components} of the relevant tensors still depend \emph{only} on $t$, with any spatial dependence relegated to the structure of the frame (i.e.\ the frame's structure constants).
	
	\section{Perturbing Bianchi spacetimes}\label{sec:perturbations}
	Since the observed signal of the CMB essentially represents a perturbation at the time of recombination (that is small fluctations from perfect homogeneity), there is value in understanding mathematically the structure of perturbations in spacetime. For FLRW spacetimes this is well-understood, see e.g.\ Peebles' seminal textbook \cite{peebles_large-scale_1980}. In terms of the Bianchi spacetimes the most popular object of study is the Bianchi I model. It is the simplest of the Bianchi models, wherein the anisotropy is captured only in temporal dependence---instead of a single scale factor as in FLRW, there are now (up to) three---and not in the choice of a non-coordinate frame; indeed, for Bianchi I this is not necessary. Its perturbations are studied in, for instance, \cite{pereira_theory_2007}. Besides Bianchi I, Bianchi VII has also received much attention previously, for it can be regarded as the anisotropic version of the flat and open FLRW models \cite{barrow_universal_1985, jaffe_evidence_2005, jaffe_bianchi_2006, pontzen_rogues_2009}. This was also investigated as a particular topic by the \emph{Planck} mission collaboration, which concluded that there is no significant evidence in favor of a Bianchi VII universe \cite{planck_collaboration_planck_2014-1, planck_collaboration_planck_2016}. Nevertheless, there does not seem to be a unifying framework for investigating perturbations in Bianchi models; our work strives towards this goal.

%	In terms of Bianchi spacetimes: although Bianchi I has received much attention, e.g.\ \cite{pereira_theory_2007}, a unifying theme is lacking in this regard. \hl{Specific attention has previously been paid to the Bianchi VII model, that might capture anistropic versions of flat and open FLRW models \cite{barrow_universal_1985}}
	
	Due to the natural frame description that we saw in \zcref{sec:theory}, Bianchi models can be treated quite similarly to FLRW in terms of the perturbation behavior. That is to say, we can follow the same procedures as we do for perturbing FLRW models, but applied to Bianchi models instead. This is the strategy we follow in \cite{scholtens_fundamental_2026}, and that we briefly present here in context. In order to capture anisotropies to the greatest extent, we assume that given some fluid flow vector field $\tb{u}$ ($\tb{u}\cdot\tb{u}=-1$), the energy-momentum tensor $\tb{T}$ of our cosmic fluid is given by \cite{ellis_relativistic_2012}
	\begin{equation}
		\tensor{T}{_\mu_\nu}=(\rho+p)\,\tensor{u}{_\mu}\tensor{u}{_\nu}+p\,\tensor{g}{_\mu_\nu}+2\,\tensor{u}{_{(\mu}}\tensor{q}{_{\nu)}}+\tensor{\pi}{_\mu_\nu},
	\end{equation}
	where $\rho$ and $p$ are, as usual, the relavistic energy density and pressure, $\tb{q}$ is the momentum density vector, and $\tb{\pi}$ the anisotropic stress tensor (with $\tensor{u}{_\alpha}\tensor{q}{^\alpha}=0$, $\tensor{\pi}{^\alpha_\alpha}=0$, and $\tensor{u}{_\alpha}\tensor{\pi}{^\mu^\alpha}=\tb{0}$). Via the Einstein field equations
	\begin{equation}
		\tensor{G}{_\mu_\nu}+\Lambda\,\tensor{g}{_\mu_\nu}=\kappa\,\tensor{T}{_\mu_\nu}
	\end{equation}
	their expressions can be derived whenever a metric and fluid flow are provided. We remind the reader that our indices refer to generally \emph{non-coordinate frames,} as introduced in \zcref{sec:theory}. The presence of the non-perfect fluid terms $\tb{q}$ and $\tb{\pi}$ is necessary so as to capture the anistropies. Given the quantities $\rho$, $p$, $\tb{q}$, and $\tb{\pi}$, it is the object of cosmological perturbation theory to determine how a perturbation in the metric $\tensor{g}{_\mu_\nu}\to\tensor{g}{_\mu_\nu}+\tensor{\delta g}{_\mu_\nu}$ leads to the perturbations of these quantities, as those represent (in principle) physical, observable quantities.
	
	For the sake of giving concrete expressions, we (i) consider only scalar perturbations, and (ii) fix our gauge to be the Newtonian gauge (but see e.g.\ \cite{nakamura_construction_2013} for more general gauge-theoretic considerations of anisotropic spacetimes). This means that we consider perturbations of the form
	\begin{equation}
		\tensor{\delta g}{_\mu_\nu}=-2(\phi+\psi)\tensor{u}{_\mu}\tensor{u}{_\nu}-2\psi\,\tensor{g}{_\mu_\nu}
	\end{equation}
	for scalar functions of spacetime $\phi$ and $\psi$. In case of the choice $\psi\equiv\phi$, which we hereafter assume, $\psi$ can be interpreted as the Newtonian potential in the weak-field approximation of general relativity---see e.g.\ \cite{wald_general_1984}. Choosing this as our metric perturbation, we have that the perturbations are given by
	\begin{subequations}\label{eq:perturbations}
		\begin{align}
			\kappa\,\delta\rho&=2\triangle\psi-2\theta\dot{\psi}+2\psi(-\tfrac{2}{3}\theta^2+2\sigma^2+6\omega^2+\Lambda+\kappa\rho)-2\kappa\,\tensor{q}{_\alpha}\tensor{\delta u}{^\alpha}. \label{eq:perturbations-density}\\ 
			3\kappa\,\delta p &=6\ddot{\psi}+2\,\tensor{\dot{u}}{^\alpha}\tensor{\psi}{_{;\alpha}}+8\theta\dot{\psi}+2\psi(4\dot{\theta}+2\theta^2+6\sigma^2+2\omega^2+\Lambda+\kappa\rho-R)-2\kappa\,\tensor{q}{_\alpha}\tensor{\delta u}{^\alpha}. \\
			\begin{split}
				\kappa\,\tensor{\delta q}{^{\langle\mu\rangle}}&=\kappa\phi\,\tensor{q}{^\mu}-\kappa\left((\rho+p)\tensor*{h}{_\alpha^\mu}+\tensor*{\pi}{_\alpha^\mu}\right)\tensor{\delta u}{^\alpha}-2\,\tensor{\dot{\psi}}{^{\langle\mu\rangle}}-\tfrac{2}{3}\theta\,\tensor{\psi}{^{|\mu}}-\tensor{4\psi}{_{|\alpha}}\tensor{\omega}{^\alpha^\mu}+\tensor{4\psi}{_{|\alpha}}\tensor{\sigma}{^\alpha^\mu} \\
				&\qquad+2\dot{\psi}\,\tensor{\dot{u}}{^\mu}-2(\phi+\psi)(\tensor{\omega}{^\alpha^\mu_{|\alpha}}+2\,\tensor{\dot{u}}{_\alpha}\tensor{\omega}{^\alpha^\mu})
			\end{split} \\
			\begin{split}
				\kappa\,\tensor{\delta\pi}{_{\langle\mu\nu\rangle}}&=-4\kappa\psi\,\tensor{\pi}{_\mu_\nu}-2\kappa\,\tensor{q}{_{\langle\mu}}\tensor{\delta u}{_{\nu\rangle}}-4\dot{\psi}\,\tensor{\sigma}{_\mu_\nu}-4\,\tensor{\dot{u}}{_{\langle\mu}}\tensor{\psi}{_{|\nu\rangle}} \\
				&\qquad-4\psi\left(\tensor{\dot{\sigma}}{_\mu_\nu}+\theta\,\tensor{\sigma}{_\mu_\nu}-2\,\tensor{\omega}{_{(\mu}^\alpha}\tensor{u}{_{\nu)}_{|\alpha}}-\tensor{\dot{u}}{_\alpha}\tensor{u}{_{(\mu}}\tensor{u}{^\alpha_{|\nu)}}+\tfrac{2}{3}\theta\,\tensor{\dot{u}}{_{(\mu}}\tensor{u}{_{\nu)}}+\tfrac{4}{3}\omega^2\,\tensor{h}{_\mu_\nu}\right)
			\end{split}
		\end{align}
	\end{subequations}
	wherein $\triangle$ is the Laplace operator on the subspace perpendicular to the fluid flow, $\Lambda$ is the cosmological constant, $R$ the scalar curvature, $\tb{\delta u}$ the perturbation in the fluid flow, and the kinematic quantities are defined according to
	\begin{equation}
		\tensor{h}{_\mu_\nu}:=\tensor{g}{_\mu_\nu}+\tensor{u}{_\mu}\tensor{u}{_\nu},\quad\theta:=\tensor{u}{^\alpha_{;\alpha}},\quad\tensor{\sigma}{_\mu_\nu}:=\tensor{u}{_{\langle\mu;\nu\rangle}},\quad\tensor{\omega}{_\mu_\nu}:=\tensor*{h}{^\alpha_\mu}\tensor*{h}{^\beta_\nu}\tensor{u}{_{[\alpha;\beta]}},\quad\tensor{\dot{u}}{^\mu}:=\tensor{u}{^\alpha}\tensor{u}{^\mu_{;\alpha}},
	\end{equation}
	with $\tensor{\sigma}{^\alpha^\beta}\tensor{\sigma}{_\alpha_\beta}=:2\sigma^2$ and $\tensor{\omega}{^\alpha^\beta}\tensor{\omega}{_\alpha_\beta}=:2\omega^2$. Importantly, due to spatial homogeneity, these quantities \emph{depend exclusively on cosmic time.} This greatly simplifies their computation and interpretation. Assuming moreover that our perturbation is isentropic (thus adiabatic) and a speed of sound $c_s^2:=\dv*{p}{\rho}$, we can combine the first two of \zcref{eq:perturbations} via $\delta p-c_s^2\,\delta\rho=0$ to obtain a single equation characterizing $\psi$:
	\begin{equation}\label{eq:HAIPE}
		\begin{split}
			\kappa(1-3c_s^2)\,\tensor{q}{_\alpha}\tensor{\delta u}{^\alpha}&=3\ddot{\psi}-3c_s^2\triangle\psi+\tensor{\dot{u}}{^\alpha}\tensor{\psi}{_{;\alpha}}+(4+3c_s^2)\theta\psi \\
			&\qquad+\psi\left(4\dot{\theta}+2(1+c_s^2)\theta^2+6(1-c_s^2)\sigma^2+2(1-9c_s^2)\omega^2+(1-3c_s^2)(\Lambda+\kappa\rho)-R\right).
		\end{split}
	\end{equation}
	This equation restricts the possible forms of the perturbation potential $\psi$ in spacetime. It is, in essence, a damped and driven wave equation, where the driving may vanish if the l.h.s.\ vanishes (e.g.\ if $\tb{q}=\tb{0}$). The key aspect of this equation is that, under the proper circumstances, it is separable into a temporal and spatial part. The former becomes a second order, linear ordinary differential equation with varying coefficients, whilst the latter reduces to the spectral analysis of a Laplacian. These are amenable to study with established numerical and analytical methods.

	\section{The CMB of a Bianchi V universe}\label{sec:bianchi-v-cmb}
\begin{figure}[t]
		\centering
		\includegraphics[width = .8\textwidth]{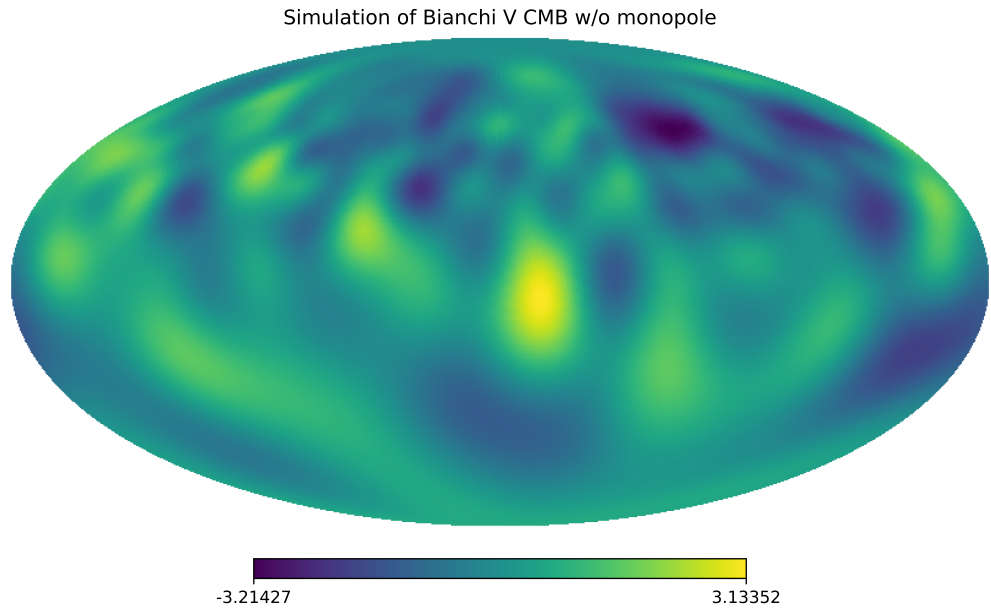}
%		\missingfigure{cmb map}
		\caption{Example map of the CMB in a Bianchi V universe, utilizing the Sachs-Wolfe effect ($v=0.85$). The units are arbitrary in this case.}
		\label{fig:cmb-map}
	\end{figure}

\begin{figure}[t]
		\centering
		\includegraphics[width = \textwidth]{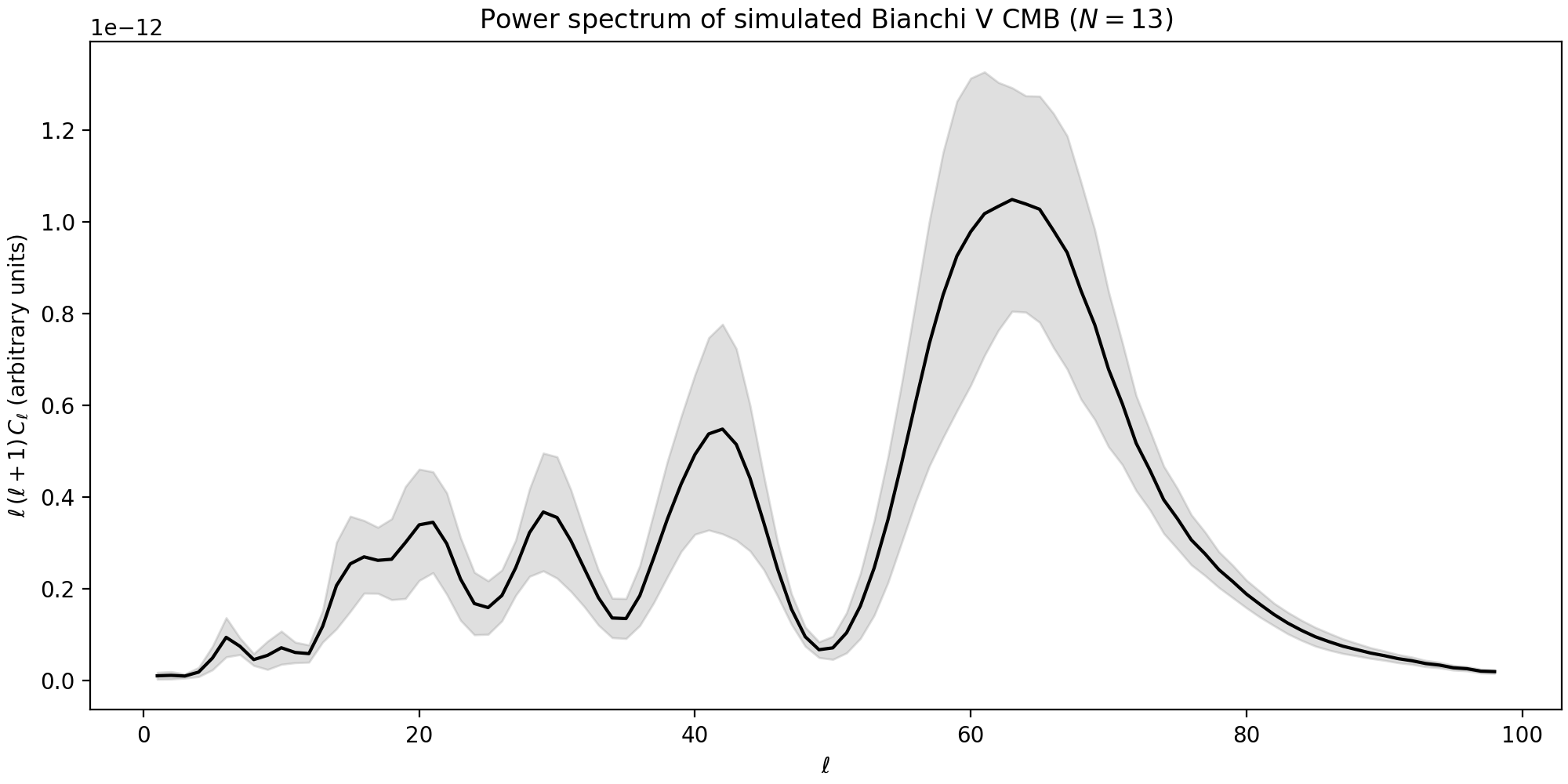}
%		\missingfigure{cmb power spectrum}
		\caption{Average power spectra of $N=13$ Bianchi V universes, with parameter $v=0.85$. The solid line represents the means of $\ell(\ell+1)C_\ell$ per $\ell$, and the shaded area the standard deviation. The units on the vertical axis are arbitrary.}
		\label{fig:cmb-power-spectra}
	\end{figure}

	In this section we will, among other things, apply \zcref{eq:HAIPE} to an example Bianchi V universe in order to simulate a CMB. Our methodology is purely based on capturing the Sachs-Wolfe effect at recombination; we do not take any other effects such as the integrated Sachs-Wolfe effect into account. In that sense, we capture purely the geometric foundation of the perturbations at the time the CMB was formed. We present a concise version of our construction, and leave the details to be found in \cite{scholtens_fundamental_2026-1}. The overall plan is as follows.
	\begin{enumerate}
		\item \emph{Set-up---choosing coordinates and finding the perturbation equation.} For our Bianchi V universe, we choose as our frame the vector fields
		\begin{equation}\label{eq:framechoice}
			\tensor{\tb{X}}{_0}=\pdv{t},\quad\tensor{\tb{X}}{_1}=-\exp(v\tensor{x}{^1})\pdv{\tensor{x}{^2}},\quad\tensor{\tb{X}}{_2}=-\exp(v\tensor{x}{^1})\pdv{\tensor{x}{^3}},\qand\tensor{\tb{X}}{_3}=-\pdv{\tensor{x}{^1}}.
		\end{equation}
		These are an explicit \emph{choice} of coordinatization, adherent to the structure of a Bianchi V universe. If another choice of coordinates is desired, for instance based on desired KVFs of the spacetime, a suitable frame can be constructed from Table 1.\ in our work \cite{scholtens_fundamental_2025}. For our fluid velocity our metric components, we choose $\tensor{u}{^\mu}=\tensor*{\delta}{^\mu_0}$ and $\tensor{g}{_\mu_\nu}=\operatorname{diag}(-1,a(t)^2,a(t)^2,a(t)^2)$, respectively. The choice of frame allows us to find the structure constants, with which we find the connection symbols in \zcref{eq:levi-civita}, and so can calculate covariant derivatives. This implies $\tb{q}=\tb{\sigma}=\tb{\omega}=\tb{\dot{u}}=\tb{0}$, $\theta=\theta(t)=3H=3\dot{a}/a$, and that the Laplacian is
		\begin{equation}
			\triangle\psi=a^{-2}(\partial_1^2+\exp(2v\tensor{x}{^1})[\partial_2^2+\partial_3^2]-2v\,\partial_1)\psi.
		\end{equation}
		Putting everything together, thus, we obtain the following perturbation equation based on \zcref{eq:HAIPE}:
		\begin{equation}
			\ddot{\psi}-c_s^2\triangle\psi+(4+3c_s^2)H\dot{\psi}+\left[2\dot{H}+3(1+c_s^2)H^2+(2+3c_s^2)a^{-2}\right]\psi=0,
		\end{equation}
		where we also used the relevant Friedmann equation $\kappa\rho+\Lambda=\tfrac{1}{3}\theta^2-3a^{-2}$ and scalar curvature $R=2\dot{\theta}+\tfrac{4}{3}\theta^2-6a^{-2}$.
		
		\item \emph{Separate the perturbations.} We notice that in this case, the perturbation equation is separable. Upon the introduction of the ansatz $\psi=T(t)S(\vb{x})$, we find that
		\begin{subequations}
			\begin{equation}
				\ddot{T}+(4+3c_s^2)H\dot{T}+\left[2\dot{H}+3(1+c_s^2)H^2+(2+(3+\lambda)c_s^2)a^{-2}\right]T=0
			\end{equation}
			as well as
			\begin{equation}
				\triangle S=-\lambda a^{-2} S,
			\end{equation}
		\end{subequations}
		where $\lambda$ is the separation constant. The former equation, being an ODE, has a guaranteed solution which can be found quickly using a computer. The latter may be separated again, leading to the solutions being a superposition of plane waves multiplied by a modified Bessel function:
		\begin{equation}\label{eq:solution_set}
			S(\vb{x})\in\left\{\cos(\tb{\mu}\cdot(\tensor{x}{^2},\tensor{x}{^3})+\phi)\exp(v\tensor{x}{^1})K_{i\nu}(v^{-1}\mu\exp(v\tensor{x}{^1}))\right\},
		\end{equation}
		where $\tb{\mu}^2=\mu^2$, $\nu=\sqrt{\lambda v^{-2}-1}>0$, $\phi$ is a free phase, and $\tb{\mu}$ is a free vector in $\mathbb{R}^2$. The constant $\mu$ was introduced as an effect of the second separation.
		
		\item \emph{Find the null geodesic ball at redshift $z=z_\text{CMB}\approx1\,100$.} Namely, this is the set of events that we observe when measuring the CMB. Since we found our connection symbols explicitly, the geodesic equation
		\begin{equation}
			\dv{\tensor{k}{^\mu}}{s}=-\tensor{\Gamma}{^\mu_\alpha_\beta}\tensor{k}{^\alpha}\tensor{k}{^\beta}
		\end{equation}
		can be solved by a computer for the given initial conditions, i.e.\ our direction of observation at the observation event. Note that these are indices in our chosen frame $\{\tensor{\tb{X}}{_\mu}\}_{\mu=0}^3$ defined in \zcref{eq:framechoice}, not the coordinate frame $\{t,\tensor{x}{^1},\tensor{x}{^2},\tensor{x}{^3}\}$! An additional step is required to trace the geodesic through coordinate space. We calculate infinitesimal redshifts following \cite[\S7.2.1]{ellis_relativistic_2012}, incrementing the redshift until the desired end point is reached. Examples of null geodesic balls in our situation for various parameter choices are given in \zcref{fig:null-geodesic-balls}.
		\begin{figure}[t]
			\centering
			\begin{subfigure}{.49\textwidth}
				\includegraphics[width = \textwidth]{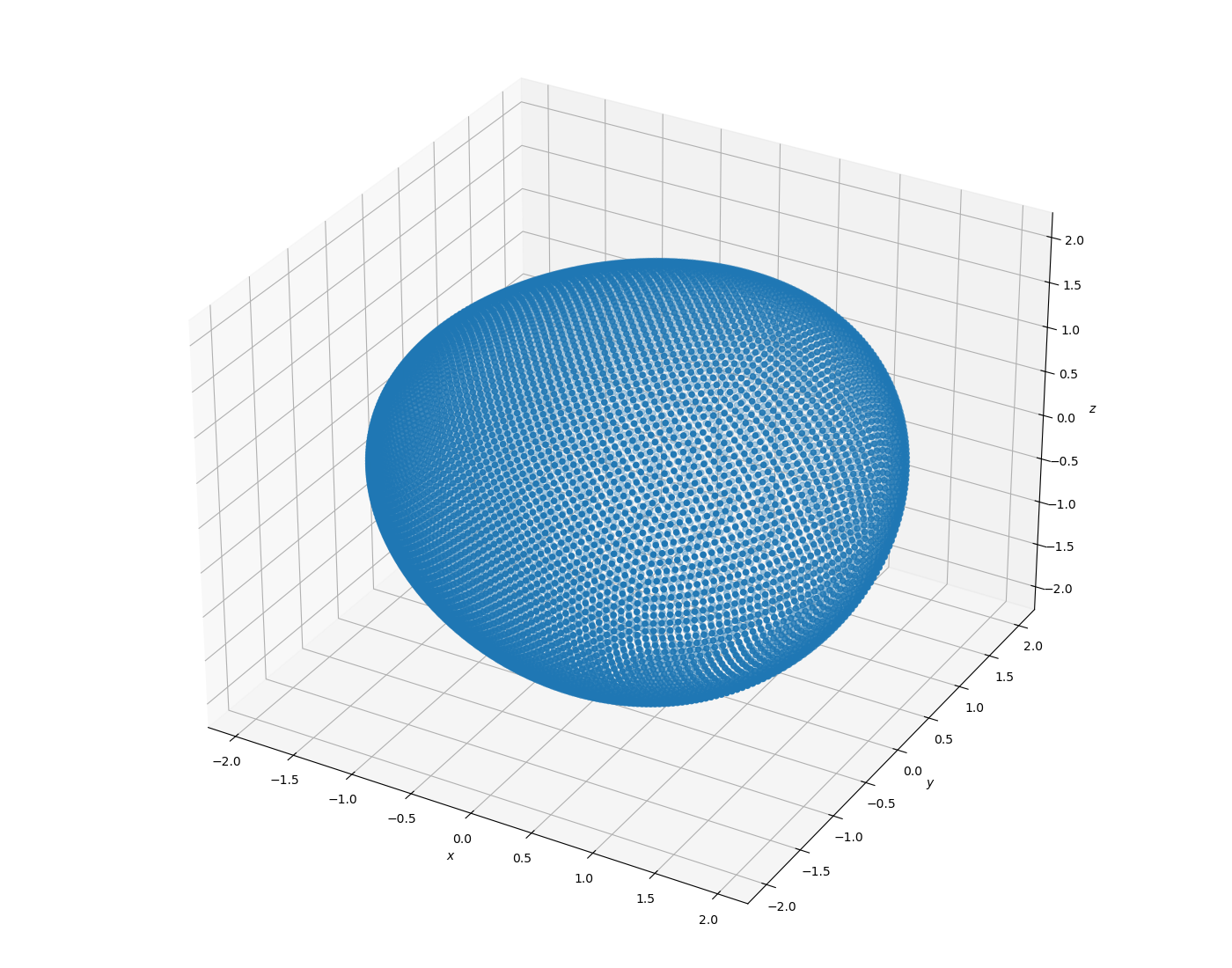}
				\caption{$v=0.15,\,\Omega_\text{mat,0}=0.98$}
			\end{subfigure}
			\begin{subfigure}{.49\textwidth}
				\includegraphics[width = \textwidth]{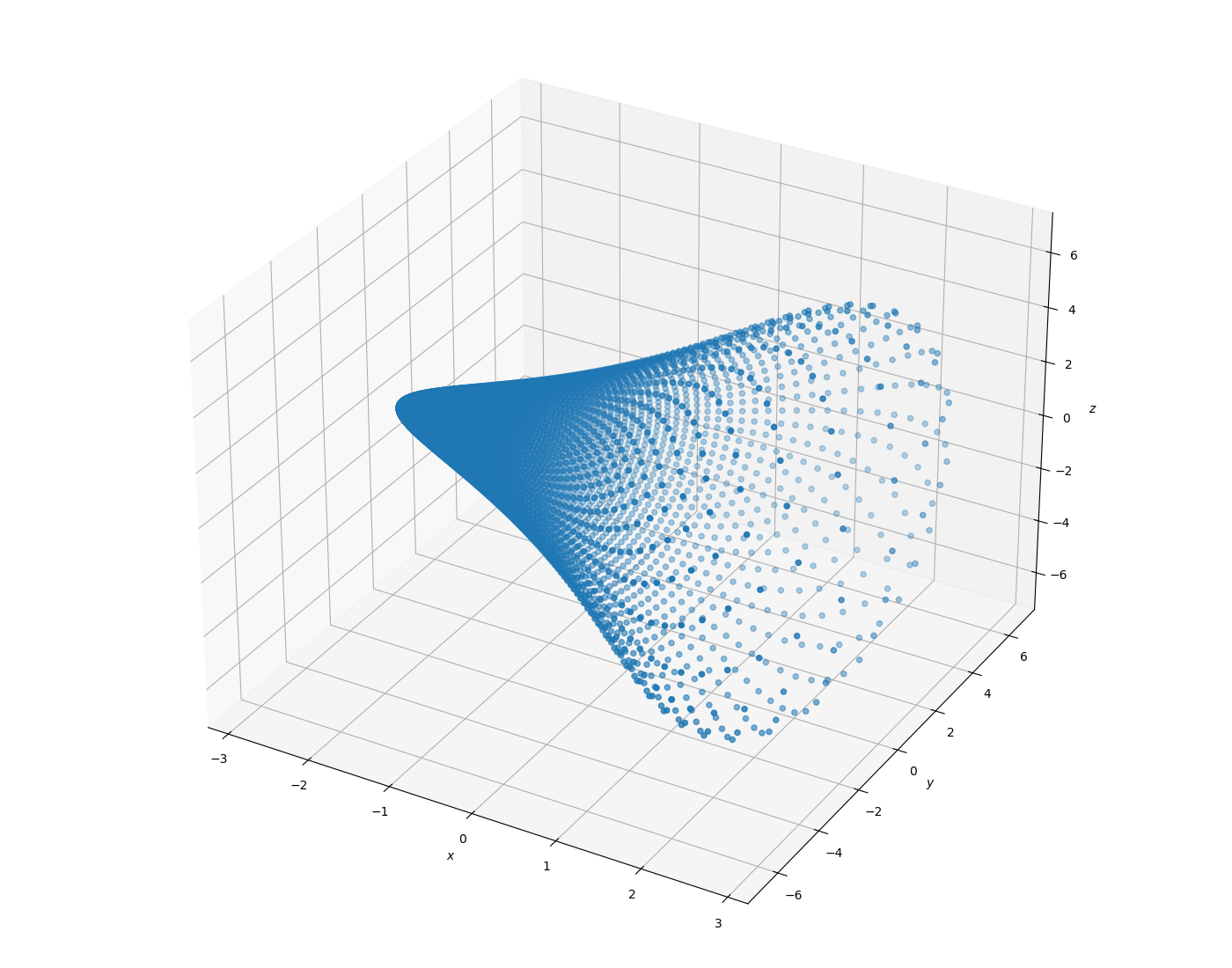}
				\caption{$v=0.85,\,\Omega_\text{mat,0}=0.28$}
				\label{fig:teardrop}
			\end{subfigure}
			\caption{Examples of null geodesic balls in our cosmology for various choices of parameters. As $v$ increases---and thus as the strength of the anisoptry increases---the shape becomes more like a teardrop. Note that $x\equiv\tensor{x}{^1}$, $y\equiv\tensor{x}{^2}$, and $z\equiv\tensor{x}{^3}$.}
			\label{fig:null-geodesic-balls}
		\end{figure}
		
		\item \emph{Superimpose all the found solutions of \zcref{eq:solution_set}, and evaluate over the null geodesic ball.} Since the perturbation equation is linear, we need to take into account a superposition of many (or all) possible solutions, which requires us to choose amplitudes and plane wave phases. Although a full physical argumentation is desirable in order to create the most physical relevance, we follow \cite{aurich_hyperbolic_2004} in choosing normally distributed amplitudes, as well as uniformly random phases.
		
		\item \emph{Retract to the observer's event to create a CMB map.} In our analysis we simply project the null geodesic ball to our night's sky, creating a CMB map. Crucially, this does not take into account effects that may disturb the radiation as it travels from its last scattering to the observer, such as the integrated Sachs-Wolfe effect. (Including such effects may the topic of future research.)
	\end{enumerate}
	At the end of this procedure, we will have produced a CMB sky map; as an example for our Bianchi V model, see \zcref{fig:cmb-map}. The notable feature is the contrast between the fluctuations on the top and lower halves of this image: the structure on the lower half seems to be more washed out than that around the top-middle section. This is due to the shape of the null geodesic ball. For this value of the anisotropy parameter, $v=0.85$, the null geodesic ball is quite tear drop-shaped---see \zcref{fig:teardrop}. This means that for smaller $\tensor{x}{^1}$, the null geodesic ball is getting more pinched, so that our observations cover less of space. In effect, we are zooming in on space with our CMB observations in the negative $\tensor{x}{^1}$-direction. The opposite conclusion holds for larger $\tensor{x}{^1}$, where the teardrop widens.
	The variation between realizations lies in the choice of plane wave phase $\phi$ and the amplitudes which we give our individual solutions. The sky map can then be processed in which ever way is deemed fit---for our purposes, we are interested in the power spectra. We use the Pyton library \texttt{healpy} in order to calculate the power spectra of the CMB sky maps \cite{zonca_healpy_2019}. We then combine these observations by finding the mean and standard deviation per $\ell$. The result of this procedure is shown in \zcref{fig:cmb-power-spectra}, with the solid line representing the mean, and the shaded area the standard deviation.
	
	\section{Discussion \& conclusion}
	In this article we have laid out the strategy that we have been developing over our papers \cite{scholtens_fundamental_2025, scholtens_fundamental_2026, scholtens_fundamental_2026-1}. We started with discussing the geometry of Bianchi models, i.e.\ spatially homogeneous yet anisotropic spacetimes, explaining that we can write such spacetimes' metrics so that the spatial dependence is captured by the frame, and the metric components capture temporal dependence---constructively for desired Killing vector fields (KVFs) through our methodology. We then conduct general relativity as usual, in that frame. Subsequently, we have written down the perturbation equations for Bianchi models in \zcref{eq:perturbations}, and combined them to one equation in \zcref{eq:HAIPE} that needs to hold for the perturbations we are considering. Finally, we used that equation in order generate CMB sky maps for a Bianchi V example, as well as calculating its power spectrum.
	
	Regarding the power spectrum shown in \zcref{fig:cmb-power-spectra}, its character is quite unlike what is known in standard literature. Although large-scale damping still holds (as we are not involving any physics beyond those at recombination), the modes for $\ell<100$ seem to fluctuate in prominence. The exact reason for why this is occurring in these spacetime models are unknown to us, as well as an appropriate interpretation of this result.
	
	Future research into this topic may center on structure formation in Bianchi universes, and in how so far it deviates from structure formation in the usual (flat) FLRW setting. This may be done by defining the density contrast as per usual, $\delta:=\delta\rho/\rho$, and studying its behavior. Particularly convenient for this end is the explicit expression of the density perturbation $\delta\rho$ as found in \zcref{eq:perturbations-density}, since it depends on $\psi$, whose behavior can be determined using \zcref{eq:HAIPE}. A downside to this approach is that, due to our choice of Newtonian gauge, there is no longer any gauge freedom to utilize, nor a manner in which to create gauge-independent conclusions. (However, one may determine the perturbation equations \emph{without} assuming a Newtonian gauge, and attempt to make progress in that fashion.)
	
	Another potential avenue is to provide a general understanding for the power spectra of CMBs in Bianchi models, potentially of all types. Such an understanding would be able to illuminate the seemingly aberrant behavior we encounter in \zcref{fig:cmb-power-spectra}. Difficulties in this regard, however, lie in a duality of complexities: for a Bianchi model, we have that both the Laplacian operator and the null geodesic balls may become (highly) no-trivial. This duality of effects, both substantially influencing CMB sky maps and their power spectra, proves quite a challenge. Additionally, if there is more than one scale factor---as a Bianchi model allows---the perturbation equation \zcref{eq:HAIPE} may cease to be separable, increasing complexity further.
	
	Progress on the above two suggestions for further research may contribute to a better understanding of anisotropic universes, and their observable signatures. In turn, that understanding will help us to gauge the validity of the cosmological principle, and, consequently, the foundations of modern cosmology as a whole.
	
%	\begin{thebibliography}{60}
%		\bibitem{Lum04} J.-P. Luminet \emph{et al}, \emph{Dodecahedral space topology as an explanation for weak wide-angle temperature correlations in the cosmic microwave background}, Nature \textbf{425} 593-595 (2004).
%	\end{thebibliography}

	\printbibliography
	
	\noindent Robbert W.\ Scholtens (contact author) \\ Kapteyn Astronomical Institute \& Bernoulli Institute for Mathematics, Computing Science, and Artificial Intelligence \\ \texttt{r.w.scholtens@rug.nl} \\
	
	\noindent Marcello Seri \\ Bernoulli Institute for Mathematics, Computing Science, and Artificial Intelligence \\ \texttt{m.seri@rug.nl} \\
	
	\noindent Holger Waalkens \\ Bernoulli Institute for Mathematics, Computing Science, and Artificial Intelligence \\ \texttt{h.waalkens@rug.nl} \\
	
	\noindent Rien van de Weygaert \\ Kapteyn Astronomical Institute \\ \texttt{weygaert@astro.rug.nl}
	
\end{document}